\documentclass[twocolumn,amsmath,amssymb]{revtex4}

\usepackage{graphicx} 
\usepackage{epstopdf}
\usepackage{graphicx}
\usepackage{dcolumn}
\usepackage{bm}
\usepackage{color}

\begin{document}

\title{Plasma effect in Silicon Charge Coupled Devices (CCDs)}

\author{J. Estrada$^{(1)}$, J. Molina$^{(2)}$, J. Blostein$^{(3)}$, G. Fern{\'{a}}ndez$^{(4)}$}

\address{$^1$Fermi National Accelerator Laboratory, Batavia, Illinois 60510, USA \\
$^2$Facultad de Ingenier{\'{\i}}a, Universidad Nacional de Asunci{\'{o}}n, Asunci{\'{o}}n, Paraguay \\
$^3$
Centro At{\'{o}}mico Bariloche and Instituto Balseiro, Comisi{\'{o}}n Nacional de Energ{\'{\i}}a At{\'{o}}mica, Universidad Nacional de Cuyo, (R8402AGP) Bariloche, Argentina \\
$^4$Universidad Nacional del Sur, Bah{\'{\i}}a Blanca, Argentina \\}

\date{\today}

\begin{abstract}
Plasma effect is observed in CCDs exposed to heavy ionizing $\alpha$-particles with energies in the range 0.5 - 5.5 MeV.
The results obtained for the size of the charge clusters reconstructed on the CCD pixels agrees with previous measurements in the high energy region ($\ge$ 3.5 MeV). The measurements were extended to lower energies using $\alpha$-particles produced by (n,$\alpha$) reactions of neutrons in a $^{10}$B target.
The effective linear charge density for the plasma column is measured as a function of energy.
The results demonstrate the potential for high position resolution in the reconstruction of alpha particles, which opens an interesting possibility for using these detectors in neutron imaging applications.
\end{abstract}

\maketitle

Scientific CCDs have been used extensively in ground- and space-based astronomy, X-ray imaging an other particle detection applications \cite{celb}. The combination of high detection efficiency, low noise, good spatial resolution, low dark current, and high charge transfer efficiency results excellent performance for detection of ionizing particles \cite{1}.

\begin{figure}
\begin{center}
\includegraphics[width=0.9\columnwidth]{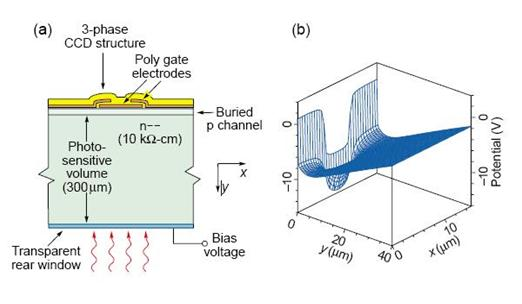}
\end{center}
\caption{(a) Cross section of the LBNL fully depleted CCD. A conventional buried channel CCD is fabricated on a high-resistivity silicon substrate. A bias voltage applied to the backside electrode results in full depletion. (b) A two-dimensional simulation showing the potential in a fully-depleted CCD that directs the photo-generated carriers into the potential wells formed by voltages on the front side CCD gate electrodes. Figure from reference \cite{proceed}}
\label{fig:DESCCD}
\end{figure}

The CCDs used in this work had been developed for the DECam wide field imager that is currently under construction \cite{DES}. The DECam CCDs detectors are devices built by Lawrence Berkeley National Laboratory (LBNL) \cite{LBNL1}, and extensively characterized at Fermilab for the DECam project \cite{2}. The DECam CCDs are 250 $\mu$m thick, fully-depleted, back-illuminated devices fabricated on high-resistivity silicon. 
The CCDs used for this study have 8 million pixels of 15 $\mu$m x 15 $\mu$m each.
Figure 1(a) shows the 3-phase, p-channel CCD design. The 10 k$\Omega$-cm resistivity corresponds to a dopant density around $10^{11}$ cm$^{-3}$, allowing a fully depleted operation at bias voltages of $\sim$ 20 - 25 V for 250 $\mu$m thick devices. The effect of the bias voltage is to remove mobile electrons from the extremely small number of phosphorous dopant atoms in the silicon, creating an electric field due to the dopant atoms that are now ionized and positively charged. The field extends essentially all the way to the backside contact, depleting the entire volume of the CCD substrate. Figure \ref{fig:DESCCD}(b) shows the modeled, 2-dimensional potential field distribution within the silicon \cite{proceed}.

When a ionizing particle penetrates the detector, it creates electron-hole pairs. Under the influence of the electric field, the holes produced near the back surface will travel the full thickness of the device to reach the potential well near the gates. During this transit inside the depletion region, a hole can also move in the direction perpendicular to the pixel boundaries causing the effect called charge diffusion.
The study of the size of X-ray hits in a back illuminated CCD provides a measurement of the lateral diffusion. 
For back illuminated detectors, a 5.9 keV X-ray will penetrate only about 20 $\mu$m into the silicon before producing ionization charge \cite{1}. The charge will travel most of the Si thickness before it can be stored under the potential well for later readout. As a result of this process, $^{55}$Fe X-rays will produce diffusion limited hits in the detector corresponding to a known energy deposition \cite{spie_estrada}. The measured charge diffusion in this condition is 2$\mu$m.

\begin{figure}
\begin{center}
\includegraphics[width=1.\columnwidth]{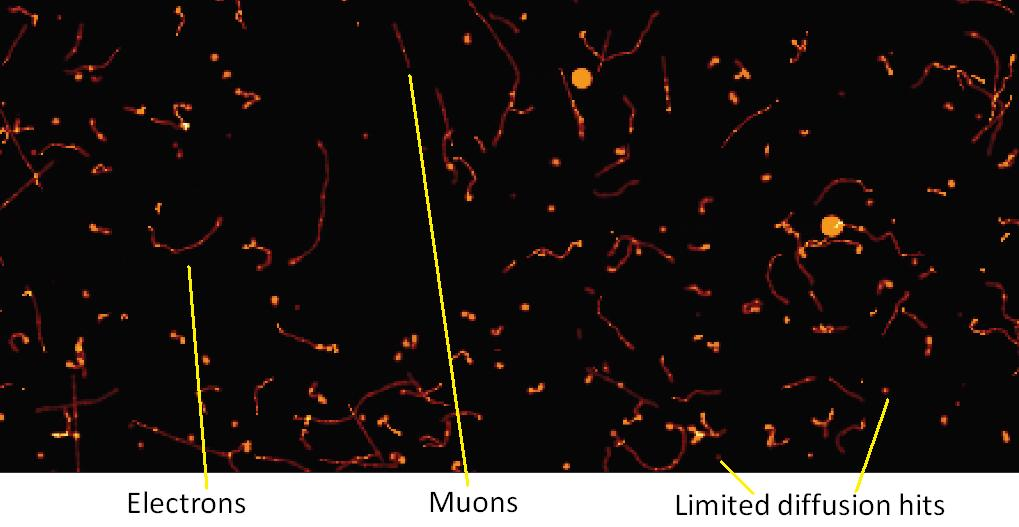}
\end{center}
\caption{Picture of a exposure of the DECam CCD to cosmic rays. Muons, electrons and low energy photons can be distinguished by their signatures on the CCD \cite{nasa}}
\label{fig:signature}
\end{figure}

More energetic particles as muons or $\beta$-particles can be distinguished from photons by the signatures that they produce in the CCD \cite{nasa}. Figure \ref{fig:signature} shows the difference among the limited diffusion hits (poinlike events), {\it{tracks}} due to muons, and {\it worms} produced by electrons.

When a heavy ionizing particle strikes the detector it creates a dense column of electron-hole pairs, which will be considered as plasma if the Debye length is small compared with the column dimensions \cite{campbell_ref4}.
In this work we study tracks from 5.5 MeV $\alpha$-particles produced in an $^{241}$Am source. 
For these $\alpha$-particles it has been shown that the charge  cloud satisfies the plasma condition (the Debye length for the charge column is $\sim $0.03 $\mu$m which is much smaller than the initial radius of the charge column $\sim$1$\mu$m \cite{medipix}). When the plasma  condition is reached there are three effects that undermine the charge  density, namely: i) erosion on the edges of the charge cloud by the presence of the external field, ii) ambipolar diffusion towards the gates of the charges inside the column,  and iii) charge recombination inside the  plasma column. The lateral size of the charge distribution measured in a pixelated detector will depend on these effects, together with the charge diffusion. Physical models of this process have been presented  in \cite{medipix, lbl89}.

The electric field created by the external voltage cannot penetrate the high density cloud of electron-hole pairs formed inside the plasma column, and the carriers will diffuse laterally towards regions of lower charge concentration in the silicon. 
This process contributes to lateral charge diffusion increasing the size of the reconstructed charge clusters. Previous work  \cite{medipix} measured this size using an $^{241}$Am $\alpha$ source, showing that the sizes of the clusters increase monotonically with the energy. In this work the measurement is extended to lower energies.

For this study two different regions were defined according to the energy of the $\alpha$-particles hitting the detectors.

\begin{figure}
\begin{center}
\includegraphics[width=1.\columnwidth]{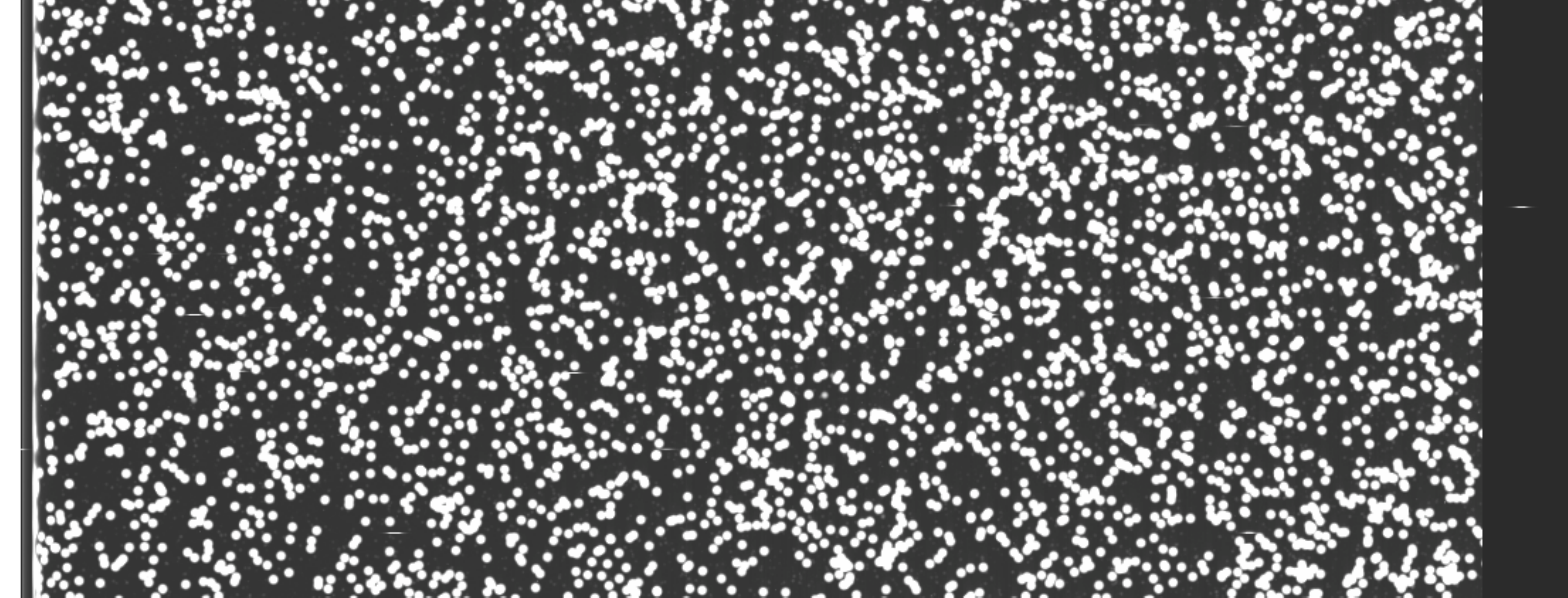}
\end{center}
\caption{Image obtained by the CCD when exposed to $^{241}$Am source}
\label{fig:alphaCCD}
\end{figure}

\begin{figure}
\begin{center}
\includegraphics[width=0.9\columnwidth]{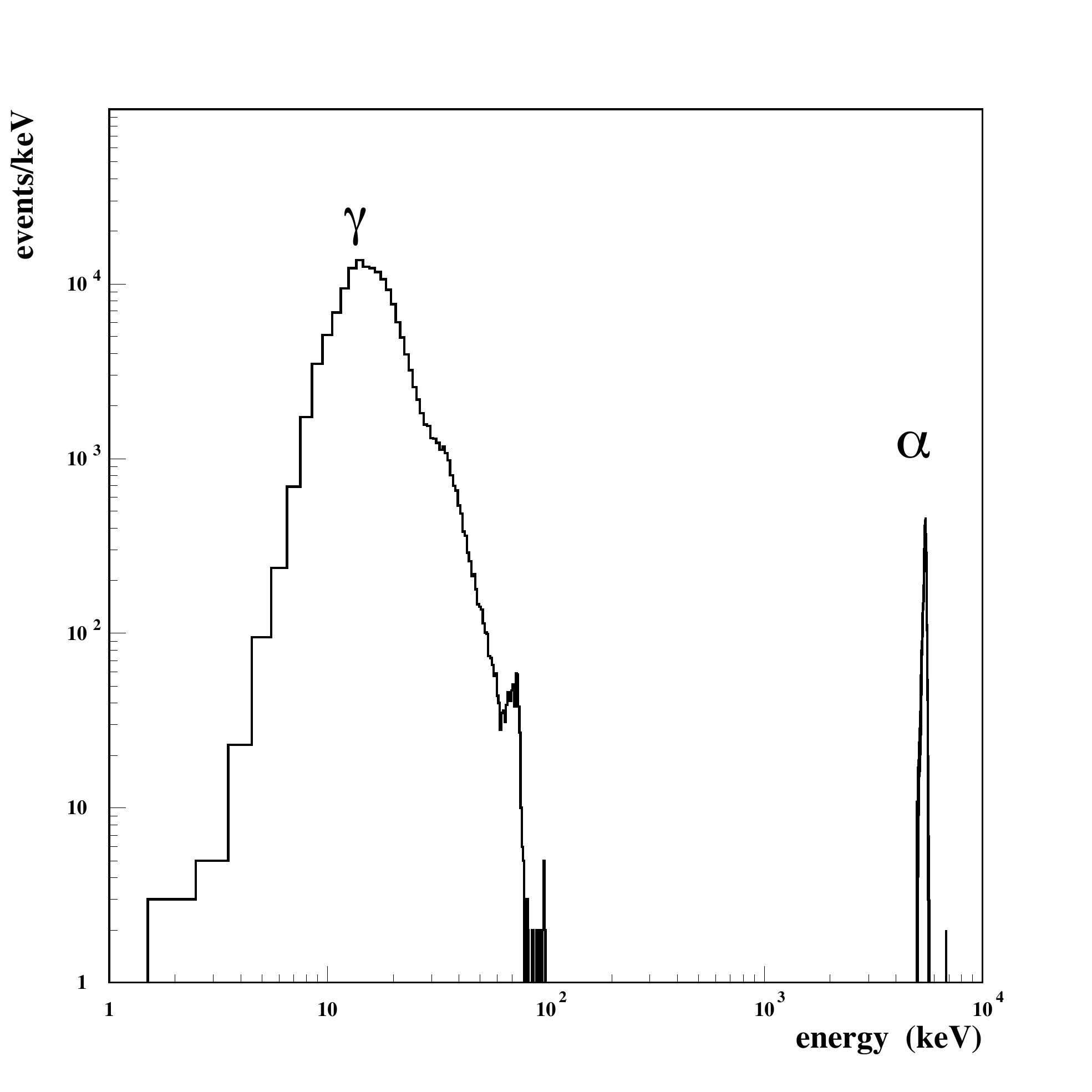}
\end{center}
\caption{Energy measured by the CCD when is exposed to a $^{241}$Am source. It can be noted the two different range in the energy of the particles detected, making a clear distinction among $\alpha$ particles and photons.}
\label{fig:alpha_spec}
\end{figure}

For the high energy region, the CCD was expose to a 0.1 $\mu$Ci $^{241}$Am source for few hours, obtaining images like the one shown in Fig \ref{fig:alphaCCD}. The corresponding energy spectrum  is shown in Fig. \ref{fig:alpha_spec}, where the 5.5 MeV peak from the $^{241}$Am can easily be distinguish from gamma signals. 

Low energy $\alpha$-particles were obtained through a nuclear reaction of the type (n, $\alpha$). 
A 5 $\mu$m thick $^{10}$B layer was placed 3mm away from the CCD, and irradiated with a $^{252}$Cf neutron source. The nuclear reactions (with their respective probabilities) inside the borated layer are: \\

n + {$^{10}$B $\rightarrow$ $^{7}$Li* (1.015MeV) + $\alpha$ (1.777MeV) $\rightarrow$  6$\%$ } \\

n + {$^{10}$B $\rightarrow$ $^{7}$Li* (840 keV) + $\alpha$ (1.470MeV) $\rightarrow   $94$\%$} \\

The particles created in this way will reach the CCD with different energies after leaving the boron layer, 
where they deposit part of their original energy.

\begin{figure}
\begin{center}
\includegraphics[width=1.\columnwidth]{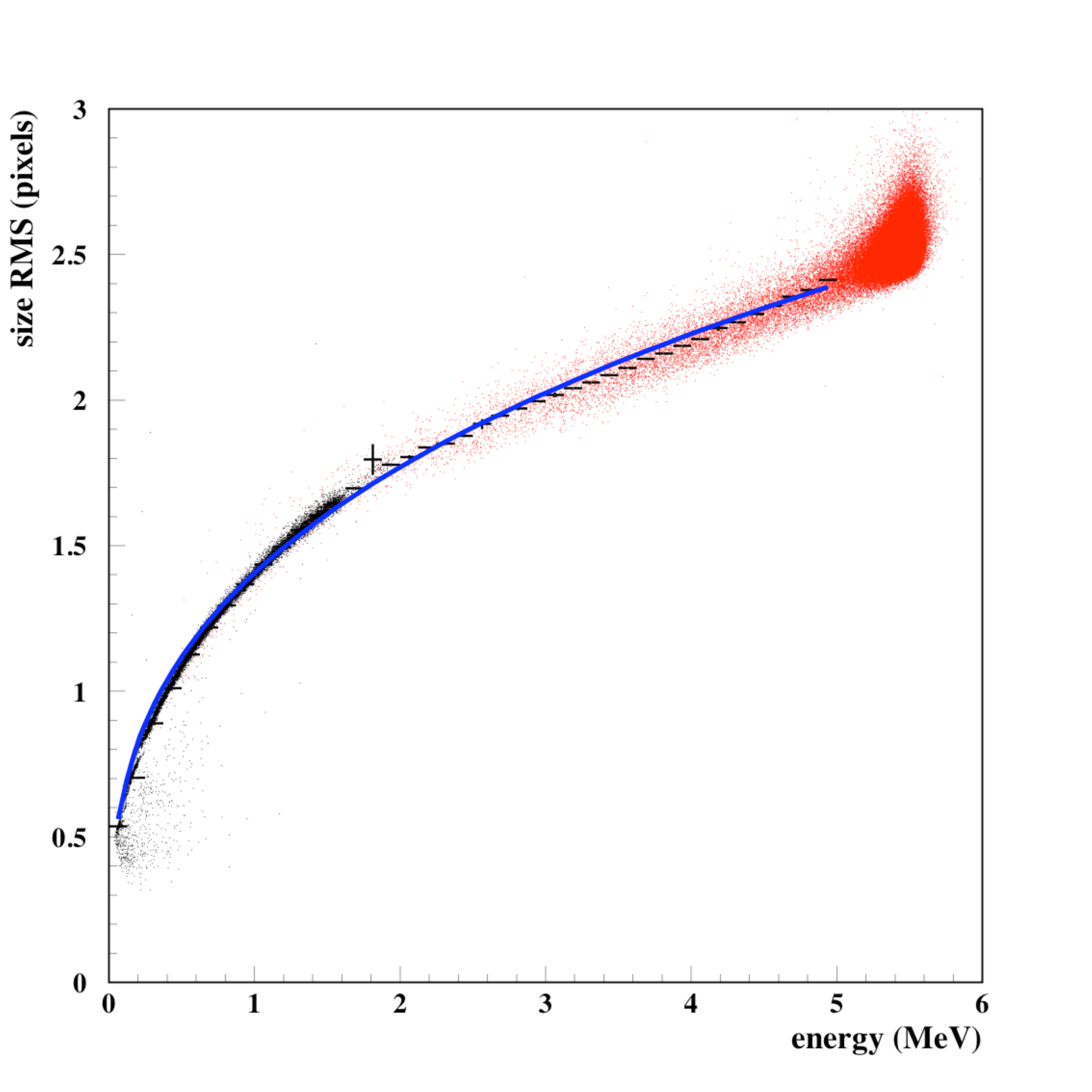}
\end{center}
\caption{Energy dependence of the cluster size for $\alpha$ particles in both regions of energy. Red poins are for the exposition of the CCD to the $^{241}$Am source, while black points are for  $\alpha$ particles coming from the (n,$\alpha$) reaction.}
\label{fig:results}
\end{figure}

 Fig. \ref{fig:results} shows the results obtained when the data from the two setups described above are put together. The red points represent events obtained with the high energy configuration, while the black points show the events obtained by the (n, $\alpha$) reaction. 

As discussed in Ref.\cite{medipix}, the expected lateral size of the reconstructed tracks is proportional to the linear density $n_{lin} $. The linear density depends on the energy ($E$) and also on the range ($R$), an can be expressed as
\begin{equation}
n_{lin}= \frac{E/k} {R(E)} \label{eq:nlin},
\end{equation}
where $k=3.6 eV/e-$ is the ionization coefficient for Silicon. The results from Fig. \ref{fig:results} can the be used for the calculation of an effective range $R_{eff}$ relevant for the plasma effect. Detailed studies have been done by other authors \cite{ultimo} to measure the range $R$ of $\alpha$-particles in silicon at these energies and the results are parameterized as 
\begin{equation}
R(E)= a\,E^b+c \label{eq:R}
\end{equation}
Where $R$ is expressed $\mu$m and $E$ in MeV. The best fit value of the parameters found in Ref.\cite{ultimo} are $a=2.13$, $b=1.45$ and $c=2.20$. $R_{\mbox{eff}}(E)$ can be parameterized in the same way. Combining Eqs. (\ref{eq:R}) and (\ref{eq:nlin}) the size $\sigma$ of the reconstructed charge clusters can be expressed as
\begin{equation}
{\sigma}^2 \propto {\Biggl{[} } {\frac{E} { a_{\mbox{eff}} \,.\,E\,^{b_{\small{\mbox{eff}}}}+c_{\mbox{eff}}}} {\Biggl{]}}^2 \label{eq:sigma}
\end{equation}
The data shown in Fig.\ref{fig:results} is fitted to Eq.(\ref{eq:sigma}) and the results are $a_{\mbox{eff}}=0.85$,  $b_{\mbox{eff}}=0.84$ and $c_{\mbox{eff}}=0$. The fitted parameters are not consistent with the measurements of range presented in Ref. \cite{ultimo}. In particular the effective range $R_{\mbox{eff}}$ calculated from our data has a different dependence with $E$, given by the 
parameter $b_{\mbox{eff}}$. This difference could be attributed to the fact that the ionization charge density is not uniform along the path of the particle, instead it follows a Bragg curve. This has also been mentioned in Ref.\cite{medipix} as a possible reason for the deviation of the model from the observed data at high fields.

The results presented here open a new interesting possibility for these type of detectors.
Due to the plasma effect, the reconstructed charge clusters for $\alpha$- particles are easily distinguishable from photons, beta-particles and muons. The reconstructed tracks produce large circular hits, with a second order moment of $\sim$2 pixels. Given the small pixels of the CCDs (15 $\mu$m x 15 $\mu$m ) this means that the centroid of each of these
$\alpha$-particle tracks can be measured with a precision of a few micrometers. Since $\alpha$-particles are commonly produced in neutron imaging applications by thermal neutron beam hitting a $^{10}$B target, the high position resolution
demonstrated in this work could help increase the spatial resolution in neutron imaging applications.

\subsection{Acknowlegments}

We thank the support of the FIUNA and CONACyT, the DECam CCD team at FNAL. Specially Kevin Kuk, Dr. Herman Cease, Greg Derylo and Andrew Lathrop.

{}

\end{document}